# Enhancing the third-order optical nonlinear performance in CMOS devices with integrated 2D graphene oxide films

David Moss

[a] Optical Sciences Centre, Swinburne University of Technology, Hawthorn, VIC 3122, Australia

**ABSTRACT**

We report enhanced nonlinear optics in complementary metal-oxide-semiconductor (CMOS) compatible photonic platforms through the use of layered two-dimensional (2D) graphene oxide (GO) films. We integrate GO films with silicon-on-insulator nanowires (SOI), high index doped silica glass (Hydex) and silicon nitride (SiN) waveguides and ring resonators, to demonstrate an enhanced optical nonlinearity including Kerr nonlinearity and four-wave mixing (FWM). The GO films are integrated using a large-area, transfer-free, layer-by-layer method while the film placement and size are controlled by photolithography. In SOI nanowires we observe a dramatic enhancement in both the Kerr nonlinearity and nonlinear figure of merit (FOM) due to the highly nonlinear GO films. Self-phase modulation (SPM) measurements show significant spectral broadening enhancement for SOI nanowires coated with patterned films of GO. The dependence of GO's Kerr nonlinearity on layer number and pulse energy shows trends of the layered GO films from 2D to quasi bulk-like behavior. The nonlinear parameter of GO coated SOI nanowires is increased 16 folds, with the nonlinear FOM increasing over 20 times to FOM > 5. We also observe an improved FWM efficiency in SiN waveguides integrated with 2D layered GO films. FWM measurements for samples with different numbers of GO layers and at different pump powers are performed, achieving up to $\approx$ 7.3 dB conversion efficiency (CE) enhancement for a uniformly coated device with 1 layer of GO and $\approx$ 9.1 dB for a patterned device with 5 layers of GO. These results reveal the strong potential of GO films to improve the nonlinear optics of silicon, Hydex and SiN photonic devices.

## 1. INTRODUCTION

All-optical integrated photonic devices are of significant interest for high-speed signal generation and processing in optical communication systems, due to the fact that they don't need the complex and inefficient optical-electrical-optical conversion [1, 2]. Triggered by a significant number of applications in telecommunications [3], metrology [4], astronomy [5], ultrafast optics [6], quantum photonics [7], and many other areas [8-10], high-performance platforms for integrated nonlinear optics has attracted much attention, and no doubt silicon-on-insulator (SOI) has led this field for several years [11-15].

While SOI has shown itself to be a leading platform for integrated photonic devices, it suffers from strong two-photon absorption (TPA) at near-infrared wavelengths, which greatly limits the nonlinear performance [2, 16], and this has motivated the use of highly nonlinear materials on chips. Other complementary metal-oxide-semiconductor (CMOS) compatible platforms including high index doped silica glass (Hydex) [17, 18] and silicon nitride (SiN) [19, 20] have a much lower TPA, but they hamper the nonlinear performance due to a comparatively low Kerr nonlinearity. To overcome these limitations, two-dimensional (2D) layered graphene oxide (GO) has received much attention among the various 2D materials due to its ease of preparation as well as the tunability of its material properties [21-29]. Previously, we reported GO films with a giant Kerr nonlinear response about 4-5 orders of magnitude higher than that of silicon and SiN [25] and demonstrated enhanced four-wave mixing (FWM) in doped silica waveguides and microring resonators (MRRs) integrated with GO films [30, 31]. Here, we demonstrate enhanced nonlinear optics in SOI nanowires [32] and SiN waveguides [33] integrated with 2D layered GO films. Owing to the strong light-matter interaction between the integrated waveguides and the highly nonlinear GO films, self-phase modulation (SPM) measurements are performed to show significant spectral broadening enhancement for SOI nanowires coated with patterned films of GO. The dependence of GO's Kerr nonlinearity on layer number and pulse energy shows interesting physical insights and trends of the layered GO films in evolving from 2D monolayers to quasi bulk-like behavior. We obtain significant enhanced nonlinear performance for the GO hybrid devices as compared with the bare waveguides, achieving the nonlinear parameter of GO-coated SOI nanowires up by 16 times, with the nonlinear figure of merit (FOM) increasing over 20 times to FOM > 5. We obtain a significant improvement in the FWM conversion efficiency (CE) of ≈ 7.3 dB for a uniformly coated SiN waveguide with 1 layer of GO and ≈ 9.1 dB for a patterned device with 5 layers of GO. These results confirm the strong potential of introducing 2D layered GO films into CMOS compatible photonic platforms to realize high-performance nonlinear photonic devices.

## 2. ENHANCED KERR NONLINEARITY IN GO-COATED SOI NANOWIRES

Figure 1a shows a schematic of an SOI nanowire waveguide integrated with a GO film. The SOI nanowire was fabricated on an SOI wafer via CMOS compatible fabrication processes, with opened windows on the silica upper cladding so as to enable GO film coating onto the SOI nanowire. The coating of 2D layered GO films was achieved by a solution-based method that yielded layer-by-layer GO film deposition. Our GO coating method can achieve precise control of the film thickness with an ultrahigh resolution of ~2 nm, which is challenging for spin coating methods. Further, our GO coating approach, unlike the sophisticated transfer processes (e.g., using scotch tape) employed for coating other 2D materials such as graphene and TMDCs [34, 35], enables transfer-free GO film coating on integrated photonic devices, with highly scalable fabrication processes as well as high fabrication stability and repeatability.

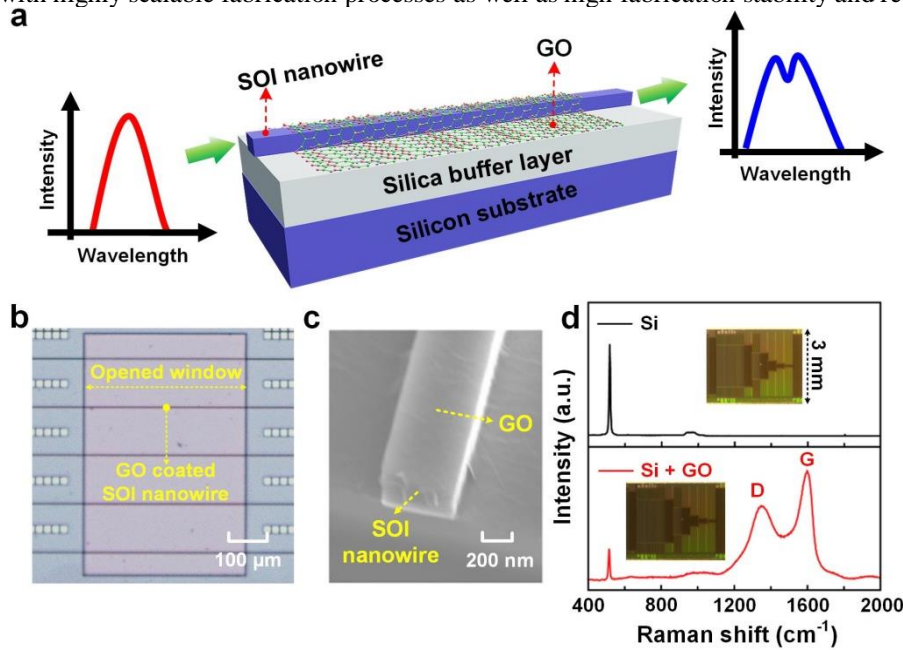

Figure 1. (a) Schematic illustration of a GO-coated SOI nanowire waveguide. (b) Microscope image of a fabricated SOI chip with a 0.4-mm-long opened window. (c) Scanning electron microscopy (SEM) image of a SOI nanowire conformally coated with 1 layer of GO. (d) Raman spectra of an SOI chip without GO and with 5 layers of GO. Insets show the corresponding microscope images.

Figure 1b shows a microscope image of a fabricated SOI chip with a 0.4-mm-long opened window. Apart from allowing precise control of the placement and coating length of the GO films that are in contact with the SOI nanowires, the opened windows also enabled us to test the performance of devices having a shorter length of GO film but with higher film thicknesses (up to 20 layers). This provided more flexibility to optimize the device performance with respect to SPM spectral broadening. Figure 1c shows the scanning electron microscopy (SEM) image of an SOI nanowire conformally coated with 1 layer of GO. Note that the conformal coating (with the GO film coated on both the top surface and sidewalls of the nanowire) is slightly different to earlier work where we reported doped silica devices with GO films only coated on the top surface of the waveguides [30, 31]. As compared with doped silica waveguides, the SOI nanowires allow much stronger light-material interaction between the evanescent field leaking from the waveguide and the GO film, which is critical to enhance nonlinear optical processes such as SPM and FWM. Figure 1d shows the successful integration of GO films which is confirmed by the representative D (1345 cm$^{-1}$) and G (1590 cm$^{-1}$) peaks of GO observed in the Raman spectrum of an SOI chip coated with 5 layers of GO. Microscope images of the same SOI chip before and after GO coating are shown in the insets, which illustrate good morphology of the films.

Figure 2 shows the results of the SPM experiments. Figure 2a-i shows the normalized spectra of the optical pulses before and after transmission through the SOI nanowires with 2.2-mm-long, 1−3 layers of GO, together with the output optical spectrum for the bare SOI nanowire, all taken with the same pulse energy of ~51.5 pJ (i.e., ~13.2 W peak power, excluding coupling loss) coupled into the SOI nanowires. As compared with the input pulse spectrum, the output spectrum after transmission through the bare SOI nanowire exhibited measurable spectral broadening. This is expected and can be attributed to the high Kerr nonlinearity of silicon. The GO-coated SOI nanowires, on the other hand, show much more significantly broadened spectra as compared with the bare SOI nanowire, clearly reflecting the improved Kerr nonlinearity of the hybrid waveguides. Figure 2a-ii shows the corresponding results for the SOI nanowires with 0.4-mm-long, 5−20 layers of GO, taken with the same coupled pulse energy as in Figure 2a-i. The SOI nanowires with a shorter GO coating length but higher film thicknesses also clearly show more significant spectral broadening as compared with the bare SOI nanowire. We also note that in either Figure 2a-i or 2a-ii, the maximum spectral broadening is achieved for a device with an intermediate number of GO layers (i.e., 2 and 10 layers of GO in a-i and a-ii, respectively). This could reflect the trade-off between the Kerr nonlinearity enhancement (which dominates for the device with a relatively short GO coating length) and loss increase (which dominates for the device with a relatively long GO coating length) for the SOI nanowires with different numbers of GO layers.

Figures 2b-i and b-ii show the power-dependent output spectra after going through the SOI nanowires with (i) 2 layers and (ii) 10 layers of GO films. We measured the output spectra at 10 different coupled pulse energies ranging from ~8.2 pJ to ~51.5 pJ (i.e., coupled peak power from ~2.1 W to ~13.2 W). As the coupled pulse energy was increased, the output spectra showed increasing spectral broadening as expected. We also note that the broadened spectra exhibited a slight asymmetry. This was a combined result of both the asymmetry of the input pulse spectrum and the free-carrier effect of silicon including both the free carrier absorption (FCA) and free carrier dispersion (FCD) [36]. Since the time response for the generation of free carries is slower compared to the pulse width, there is a delayed impact of FCA on the pulse shape, which leads the spectral asymmetry of the optical pulses. The FCD further broadens the asymmetry induced by FCA, resulting in more obvious spectral asymmetry at the output.

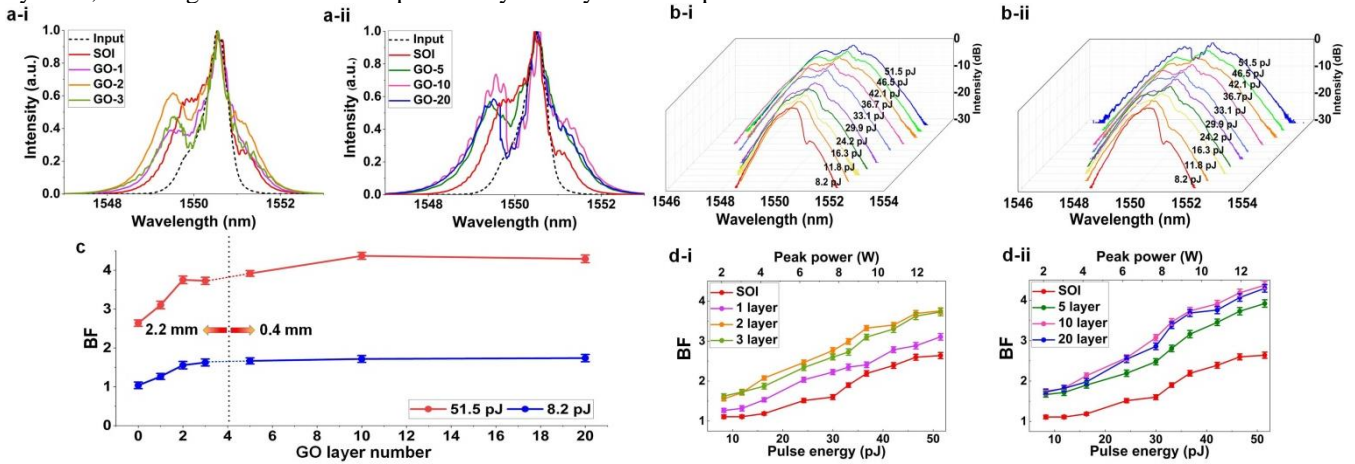

Figure 2. SPM experimental results. (a) Normalized spectra of optical pulses before and after going through the GO-coated SOI nanowires at a coupled pulse energy of ~51.5 pJ. (b) Optical spectra measured at different pulse energies for the GO-coated SOI nanowires. (c) BFs of the measured output spectra versus GO layer number at fixed coupled pulse energies of 8.2 pJ and 51.5 pJ. (d) BFs of the measured output spectra versus coupled pulse energy (or coupled peak power). In (a), (b) and (d), (i) and (ii) show the results for the SOI nanowires with 2.2-mm-long, 1−3 layers of GO and with 0.4-mm-long, 5−20 layers of GO, respectively. In (a), (c) and (d), the corresponding results for the bare SOI nanowires are also shown for comparison.

To quantitively analyze the spectral broadening of the output spectra, we introduce the concept of a broadening factor (BF, defined as the square of the pulse'rms spectral width at the waveguide output facet divided by the corresponding value at the input [37]). Figure 2c shows the BFs of the measured output spectra after transmission through the bare SOI nanowire and the GO-coated SOI nanowires at pulse energies of 8.2 pJ and 51.5 pJ. The GO-coated SOI nanowires show higher BFs than the bare SOI nanowires (i.e., GO layer number = 0), and the BFs at a coupled pulse energy of 51.5 pJ are higher than those at 8.2 pJ, agreeing with the results in Figures 2a and 2b, respectively. At 51.5 pJ, BFs of up to 3.75 and 4.34 are achieved for the SOI nanowires with 2 and 10 layers of GO, respectively. This also agrees with the results in Figure 2a − with the maximum spectral broadening being achieved for an intermediate number of GO layers due to the trade-off between the Kerr nonlinearity enhancement and increase in loss. The BFs of the output spectra versus coupled pulse energy are shown in Figures 2d-i and 2d-ii for the SOI nanowires with 1−3 layers and 5−20 layers of GO, respectively. The BFs increase with coupled pulse energy, reflecting a more significant spectral broadening that agrees with the results in Figure 2b.

Figure 3a shows Kerr coefficient ($n_2$) of the GO films versus layer number for fixed coupled pulse energies of 8.2 pJ and 51.5 pJ, which is extract from the effective nonlinear parameter ($\gamma_{eff}$) of the hybrid waveguides using the following equation [30]:

$$\gamma_{eff} = \frac{2\pi}{\lambda_c} \frac{\iint_D n_0^2(x,y) n_2(x,y) S_z^2 \, dxdy}{\left[\iint_D n_0(x,y) S_z \, dxdy\right]^2} \tag{1}$$

where $\lambda_c$ is the pulse central wavelength, $D$ is the integral of the optical fields over the material regions, $S_z$ is the time-averaged Poynting vector calculated using Lumerical FDTD commercial mode solving software, $n_0(x, y)$ and $n_2(x, y)$ are the linear refractive index and $n_2$ profiles over the waveguide cross section, respectively.

The picosecond optical pulses used in our experiment had a relatively small spectral width (< 10 nm), we therefore neglected any variation in $n_2$ arising from its dispersion and used $n_2$ instead of the more general third-order nonlinearity $\chi^{(3)}$ in our subsequent analysis and discussion. The values of $n_2$ are over three orders of magnitude higher than that of silicon and agree reasonably well with our previous waveguide FWM [30] and Z-scan measurements [28]. Note that the layer-by-layer characterization of $n_2$ for GO is challenging for Z-scan measurements due to the weak response of extremely thin 2D films [25, 28]. The high $n_2$ of GO films highlights their strong Kerr nonlinearity for not only SPM but also other third-order ($\chi^{(3)}$) nonlinear processes such as FWM, and possibly even enhancing ($\chi^{(3)}$) for third harmonic generation (THG) and stimulated Raman scattering [38-40]. In Figure 3a, $n_2$ (both at 51.5 pJ and 8.2 pJ) decreases with GO layer number, showing a similar trend to WS$_2$ measured by a spatial-light system [41]. This is probably due to increased inhomogeneous defects within the GO layers as well as imperfect contact between the different GO layers. Although the $n_2$ of GO decreases with layer number, the increase in mode overlap with GO more than compensates for this, resulting in a net increase in $\gamma_{eff}$ with layer number. At 51.5 pJ, $n_2$ is slightly higher than at 8.2 pJ, indicating a more significant change in the GO optical properties with inceasing power. We also note that the decrease in $n_2$ with GO layer number becomes more gradual for thicker GO films, possibly reflecting the transition of the GO film properties towards bulk material properties − with a thickness independent $n_2$.

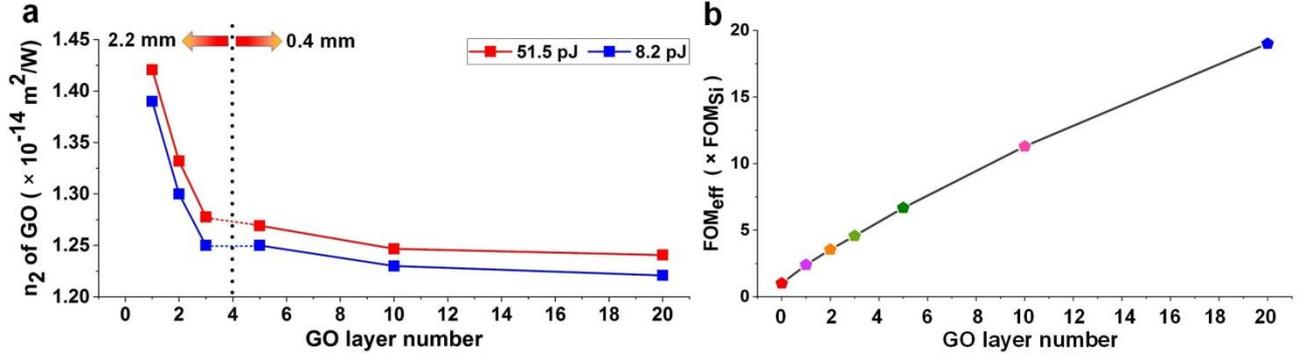

Figure 3. (a) $n_2$ of GO versus layer number at fixed coupled pulse energies of 8.2 pJ and 51.5 pJ. (b) $FOM_{eff}$ of the hybrid waveguides versus GO layer number. The coupled pulse energy is ~51.5 pJ. and the corresponding results for silicon (GO layer number = 0) is also shown for comparison.

To quantitively analyze the improvement in the nonlinear performance of the GO-coated SOI nanowires, we further calculated the effective nonlinear FOM ($FOM_{eff}$) for the GO-coated SOI nanowires. The resulting $FOM_{eff}$ (normalized to the FOM of silicon) is shown in Figure 3b where we see that a very high $FOM_{eff}$ of 20 times that of silicon is achieved for the hybrid SOI nanowires with 20 layers of GO. This is remarkable since it indicates that by coating SOI nanowires with GO films, not only can the nonlinearity be significantly enhanced but the relative effect of nonlinear absorption can be greatly reduced as well. This is interesting given that the GO films themselves cannot be described by a nonlinear FOM since the nonlinear absorption displays saturable absorption (SA) rather than TPA, and yet nonetheless the GO films still are able to reduce the effective TPA coefficient ($\beta_{TPA, eff}$) of the hybrid waveguides, thus improving the overall nonlinear performance.

## 3. ENHANCED FWM IN GO-COATED SIN WAVEGUIDES

Figure 4a shows the SiN waveguide integrated with a GO film, along with a schematic showing atomic structure of GO with different oxygen functional groups (OFGs) such as hydroxyl, epoxide and carboxylic groups. SiN waveguides with a cross section of 1.6 μm × 0.66 μm were fabricated via annealing-free and crack-free processes that are compatible with CMOS fabrication [37, 42]. Layered GO films were coated on the top surface of the chip by a solution-based method that yielded layer-by-layer film deposition, as mentioned in section 2 [29, 30, 43]. Figure 4b shows a microscope image of a SiN waveguide patterned with 10 layers of GO, which illustrates the high transmittance and good morphology of the GO films.

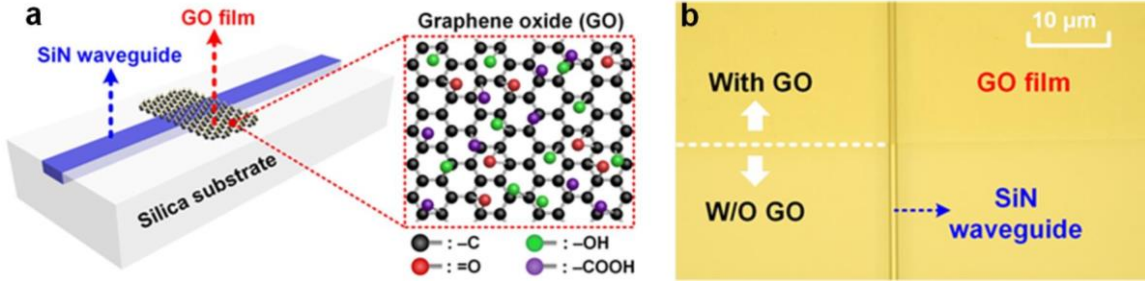

Figure 4. (a) Schematic illustration of GO-coated SiN waveguide. Inset shows the schematic atomic structure of GO. (b) Microscope image of a SiN waveguide patterned with 10 layers of GO.

Figure 5 shows the experimental FWM optical spectra for the SiN waveguides uniformly coated with 1 and 2 layers of GO (Figure 5a-i) together with the FWM spectrum of the bare SiN waveguide. For comparison, we kept the same power of 23 dBm for both the pump and signal before the input of the waveguides, which corresponded to 18 dBm power for each coupled into the waveguides. The difference among the baselines of the spectra reflects the difference in waveguide propagation loss for different samples. It can be seen that although the hybrid waveguide with 1 layer of GO film had an additional propagation loss of ≈7.1 dB, it clearly shows enhanced idler output powers as compared with the bare SiN waveguide. The CE (defined as the ratio of the output power of the idler to the input power of the signal, i.e., Pout, idler / Pin, signal) of the SiN waveguides without GO and with 1 layer of GO were ≈-65.7 dB and ≈-58.4 dB, respectively, corresponding to a CE enhancement of ≈7.3 dB for the hybrid waveguide. In contrast to the positive CE enhancement for the hybrid waveguide with 1 layer of GO, the change in CE for the hybrid waveguide with 2 layers of GO was negative. This mainly resulted from the increase in propagation loss with GO layer numbers.

Figure 5a-ii shows the FWM spectra of the SiN waveguides with 5 and 10 layers of patterned GO films. The coupled continuous-wave (CW) pump and signal power (18 dBm for each) was the same as that in Figure 5a-i. The SiN waveguides with patterned GO films also had an additional insertion loss as compared with the bare waveguide, while the results for both 5 and 10 GO layers show enhanced idler output powers. In particular, there is a maximum CE enhancement of ≈ 9.1 dB for the SiN waveguide patterned with 5 layers of GO, which is even higher than that for the uniformly coated waveguide with 1 layer of GO. This reflects the trade-off between FWM enhancement (which dominates for the patterned devices with a short GO coating length) and loss (which dominates for the uniformly coated waveguides with a much longer GO coating length) in the GO-coated SiN waveguides.

Figure 5b shows the measured CE versus pump power for the uniformly coated and patterned devices, respectively. The plots show the average of three measurements on the same samples and the error bars reflect the variations, showing that the measured CE is repeatable. As the pump power was increased, the measured CE increased linearly with no obvious saturation for the bare SiN waveguide and all the hybrid waveguides, indicating the low TPA of both the SiN waveguides and the GO films. For the bare waveguide, the dependence of CE versus pump power shows a nearly linear relationship, with a slope rate of about 2 for the curve as expected from classical FWM theory [44-54]. For the GO-coated waveguides, the measured CE curves have shown slight deviations from the linear relationship with a slope rate of 2, particularly at high light powers. Figure 5c compares the CE of the hybrid waveguides with four different numbers of GO layers (i.e., 1, 2, 5, 10), where we see that the hybrid waveguide with an intermediate number of GO layers has the maximum CE. This reflects the trade-off between $\gamma$ and loss in the hybrid waveguides, which both increase with GO layer number.

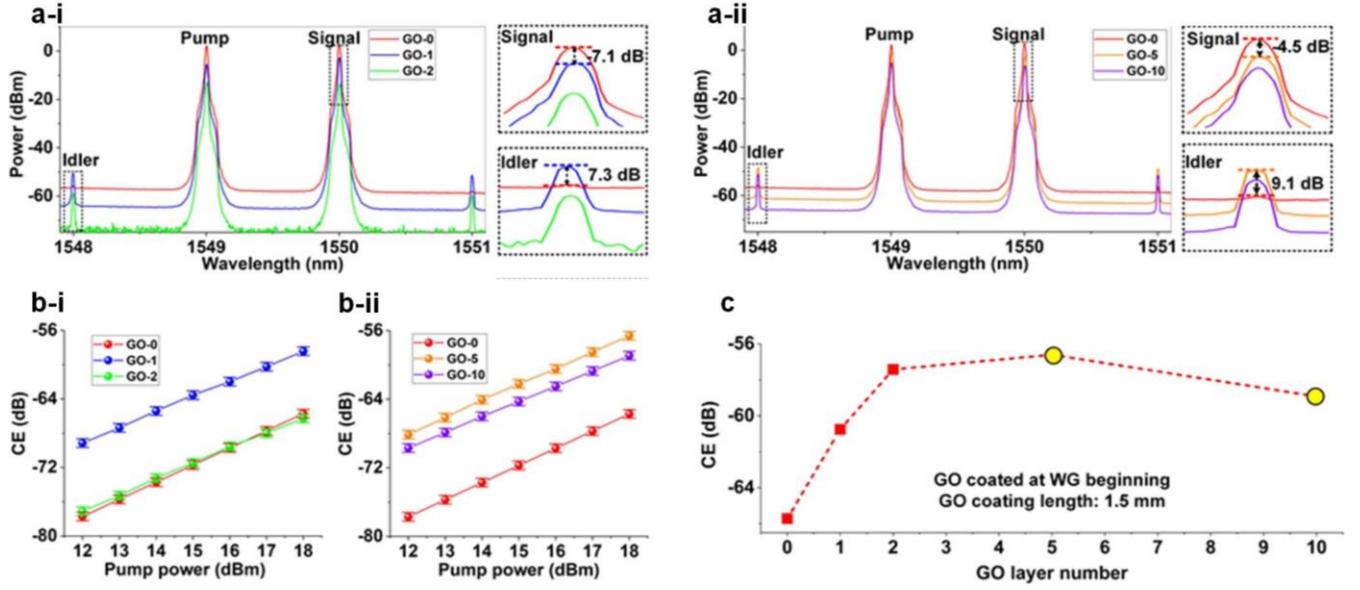

Figure 5. FWM experimental results. (a) FWM optical spectra. Insets show zoom-in view around the signal and idler. (b) CE versus pump power coupled into the waveguides. In (a) and (b), (i) shows the results for SiN waveguides uniformly coated with 1 and 2 layers of GO and (ii) shows the results for SiN waveguides patterned with 5 and 10 layers of GO. (c) Calculated CE as functions of GO layer number. In (c), the corresponding results for the bare SiN waveguide (GO-0) are also shown for comparison. The coating length is 1.5 mm and the GO coating position is at waveguide beginning.

Table 1 compares the performance of SOI nanowires, SiN, and Hydex waveguides incorporated with GO films. As we can see in the Table, the dimensions of the three CMOS compatible photonic platforms were quite different. The SOI nanowire had the smallest waveguide dimensions and the tightest mode confinement, resulting in significantly increased mode overlap with the GO film. This resulted in a significantly increased nonlinear parameter $\gamma$, but also the largest excess propagation loss induced by the GO film. Mode overlap is a key factor for optimizing the trade-off between the Kerr nonlinearity and loss when introducing 2D layered GO films onto different integrated platforms to enhance the nonlinear optical performance.

Table 1. Performance comparison of SOI nanowire, SiN, and Hydex waveguides integrated with 2D layered GO films. WG: waveguide

|  | Waveguide dimension (μm) | $EPL_{GO-1}$ [a] (dB/cm) | $\gamma_{hybrid}$ [b] ($W^{-1}m^{-1}$) | $n_2$ of GO ($m^2/W$) [c] |
|---|---|---|---|---|
| SOI-GO hybrid WG | Width: 0.5 Height: 0.22 | 20.5 | GO-1: 668, GO-10: 2905 | $1.22 \times 10^{-14} \sim 1.42 \times 10^{-14}$ |
| SiN-GO hybrid WG | Width: 1.60 Height: 0.66 | 3.1 | GO-1: 13.14, GO-10: 167.14 | $1.28 \times 10^{-14} \sim 1.41 \times 10^{-14}$ |
| Hydex-GO hybrid WG | Width: 2 Height: 1.5 | 1 | GO-1: 0.61, GO-10: n/a [d] | $1.5 \times 10^{-14}$ |

a) $EPL_{GO-1}$: excess propagation loss induced by GO for the hybrid waveguide with 1 layer of GO.
b) $\gamma_{hybrid}$: nonlinear parameter of the hybrid waveguide with 1 layer (GO-1) and 10 (GO-10) layers of GO.
c) The $n_2$ values of GO are extracted from the nonlinear SPM or FWM experiments.
d) Only hybrid waveguides with 1-5 layers of GO were characterized.

## 4. CONCLUSION

We demonstrate enhanced nonlinear optics including Kerr nonlinearity and FWM in SOI nanowires, Hydex and SiN waveguides and ring resonators incorporated with layered GO films. We achieve precise control of the placement, thickness, and length of the GO films using layer-by-layer coating of GO films followed by photolithography and lift-off. Owing to the strong mode overlap between the platforms and the highly nonlinear GO films, we achieve a high nonlinear parameter of GO coated SOI nanowires up to 16 times and an improved nonlinear FOM of up to a factor of 20. We obtain a significant improvement in the FWM CE of ≈ 7.3 dB for a uniformly coated SiN waveguide with 1 layer of GO and ≈ 9.1 dB for a patterned device with 5 layers of GO. These results verify the enhanced nonlinear optical performance of silicon, Hydex and SiN photonic devices achievable by incorporating 2D layered GO films.

Competing interests: The authors declare no competing interests.